\magnification1200
\font\BBig=cmr10 scaled\magstep2
\font\BBBig=cmr10 scaled\magstep3


\def\title{
{\bf\BBBig
\centerline{Vortex solutions in axial or chiral}
\bigskip
\centerline{coupled}
\bigskip
\centerline{non-relativistic spinor- Chern-Simons theory}
}}

\def\foot#1{
\footnote{($^{\the\foo}$)}{#1}\advance\foo by 1
} 
\def\ccr{\cr\noalign{\medskip}}


\def\authors{
\centerline{
Z. N\'EMETH\foot{Institute for Theoretical Physics,
E\"otv\"os University, H-1088 BUDAPEST,
\hfill\break
Puskin u. 5-7 (Hungary). e-mail: nemethz@ludens.elte.hu}
}}

\def\runningauthors{N\'emeth}

\def\runningtitle{Spinor vortices in axial\dots}


\voffset = 1cm 
\baselineskip = 14pt 

\headline ={
\ifnum\pageno=1\hfill
\else\ifodd\pageno\hfil\tenit\runningtitle\hfil\tenrm\folio
\else\tenrm\folio\hfil\tenit\runningauthors\hfil
\fi
\fi}

\nopagenumbers
\footline={\hfil} 


\def\and{\qquad\hbox{and}\qquad}

\def\smallcirc{{\raise 0.5pt \hbox{$\scriptstyle\circ$}}} 
\def\smallover#1/#2{\hbox{$\textstyle{#1\over#2}$}} %
\def\2{{\smallover 1/2}}
\def\ccr{\cr\noalign{\medskip}}
\def\parag{\hfil\break} 
\def\={\!=\!}
\def\D{{D\mkern-2mu\llap{{\raise+0.5pt\hbox{\big/}}}\mkern+2mu}\ }

\def\vx{\vec{x}}
\def\L{{\cal L}}
\def\vsigma{\vec{\sigma}}


\newcount\ch 
\newcount\eq 
\newcount\foo 
\newcount\ref 

\def\chapter#1{
\parag\eq = 1\advance\ch by 1{\bf\the\ch.\enskip#1}
}

\def\equation{
\eqno(\the\ch.\the\eq)\global\advance\eq by 1
}
\def\reference{
\parag [\number\ref]\ \advance\ref by 1
}

\ch = 0 
\foo = 1 
\ref = 1 


\title
\vskip 1.5cm
\authors
\vskip .25in

\parag
{\bf Abstract.}

{\sl 
The interaction of a spin 1/2 particle 
(described by the non-relativistic "Dirac" 
equation of L\'evy-Leblond) with Chern-Simons gauge fields is studied. It 
is shown, that similarly to the four dimensional spinor models,
there is a consistent possibility of coupling them also by axial or chiral
type currents. Static self dual
vortex solutions together with a 
vortex-lattice are found with the new couplings}.
\bigskip
%
\noindent{
(\the\day/\the\month/\the\year)}
\bigskip

\vfill\eject
\chapter{Introduction}

Several exact static solutions of the field equations have been found 
in models, describing the coupling of 
Chern-Simons (CS) gauge fields with various types of  
matter fields. 
Among them is 
the well known self-dual solution of Jackiw and Pi [1] in a model of CS gauge
fields and non relativistic matter field, described by the nonlinear 
Schr\"odinger equation. Another family of solutions of this 
model, obeying periodic boundary conditions and
forming a vortex lattice on the plane, is presented by Olesen [2]. 
Duval Horv\'athy and Palla [3] described 
a system of nonrelativistic spinors coupled to CS fields,  
where the matter field satisfies the 
L\'evy-Leblond equations [3,4], and they also presented some 
static self-dual solutions. 

In this paper the nonrelativistic spinor field is also assumed to  
satisfy the L\'evy-Leblond equations, but
the current, coupling it to the CS fields, is not the vector current, but is  
a different, \lq axial' or \lq chiral type' one. 
The \lq axial type' current is the result of a light-like Kaluza-Klein 
reduction from the 3+1 dimensional axial current 
in a similar way as the current in [3] is the reduced form of the 
3+1 dimensional vector current.
From the 2+1 dimensional point of wiev, with this coupling, the 
gauge field would be 
coupled to the spin density of the matter. The chiral currents are linear 
combinations of the vector and axial type ones. Various models with these 
couplings are studied and 
it is shown, that even with the new couplings the system 
of field equations is explicitly 
solvable (at least in the self dual case) and several
exact solutions are presented.

The paper is organized as follows: 
In Sec. 2. the axially coupled system is described, 
and in Sec. 3. I solve this system with   
the self-duality condition used in [3].
In Sec. 4. a new self-duality condition is put forward, allowing more general
solutions, and is shown how this new condition leads to the 
vortex lattice solutions. 
Sec. 5. deals with the chiral coupling.

\chapter{The model}

        In this paper we are interested in a model described by the 
Levi-Leblond and Chern-Simons field equations, like in the
nonrelativistic spinor Chern-Simons theory [3],   
but coupled with an axial type 
current.
        The Levi-Leblond equations (the 2+1 dimensional version of the non
relativistic \lq Dirac equation'), eq. (2.1), and the field current identity 
(FCI), eq. (2.2), can be written as:
$$\left\{
\matrix{
(\vec{\sigma}\cdot\vec{D})\,\Phi
&+\hfill&2m\,\chi&=&0,
\ccr
D_t\,\Phi&+\hfill&i(\vec{\sigma}\cdot\vec{D})\,\chi\hfill&=&0,
\cr}\right.
\equation
$$
$$\left\{
\matrix{
\kappa B\equiv\kappa\epsilon^{ij}\partial_iA^j=
-e\varrho,
\ccr
\kappa E^i\equiv \kappa(-\partial_iA^0-\partial_tA^i)=
e\,\epsilon^{ij}J^j,
\cr}\right.
\equation
$$
where 
$$
\Phi=\pmatrix{\Psi_+\cr\Psi_-\cr}
\and
\chi=\pmatrix{\chi_+\cr\chi_-\cr},
\equation
$$
are two-component `Pauli' spinors and 
$(\vec{\sigma}\cdot\vec{D})=\sum\limits_{j=1}^2\sigma^jD_j$, 
with $D_j=\partial_j-ie\,A_j$, $\sigma^j$ denoting the Pauli matrices.
Here $\kappa$ stands for the CS self coupling, $e$ is the gauge coupling and
$J^\alpha\=(\rho,\vec{J})$ denote the charge and current densities that couple
the LL and CS equations. In the model of Duval, Horv\`athy and Palla 
$$
\varrho=|\Phi|^2,
\and
\vec{J}=i\big(\Phi^\dagger\vec{\sigma}\,\chi
-\chi^\dagger\vec{\sigma}\,\Phi\big),
\equation
$$
while in the new model these quantities are given as
$$
\varrho=|\Psi_+|^2-|\Psi_-|^2,
\and
\vec{J}=-i\big(\Phi^\dagger\vec{\sigma}\sigma_3\,\chi
+\chi^\dagger\vec{\sigma}\sigma_3\,\Phi\big).
\equation
$$  

The FCI is self consistent if the continuity equation holds for $J^\alpha$.
It can be seen easily that it is satisfied also for the new current, 
thus the system is self consistent. 

        Let us see why this current is of axial type!
        The LL equation with the 2+1 dimensional FCI can be obtained in a 
Kaluza-Klein type reduction procedure from the higher (i.e. 3+1) dimensonal 
system of massless Dirac equation and 
3+1 dimensional form of FCI [5]. ( It is a 
reduction from a four dimensional Bargmann manifold $(M,g,\xi)$ in
a trivializing coordinate  system $(t,x,y,s)$,
where the particular vector field $\xi$ is given by $\xi=\partial_s$, to the 
quotient, $Q$, of $M$, by the flow of $\xi$, in coordinates:
$t$ and $\vx=(x,y)$.) The four dimensional system has the form:
$$\left\{
\matrix{\D\psi=0,
\qquad 
\qquad
D_\xi\psi=im\psi,
\ccr
\kappa f_{\mu\nu}=
e\sqrt{-g}\,\epsilon_{\mu\nu\rho\sigma}\,\xi^\rho j^\sigma,
\qquad 
\qquad 
\partial_{[\mu}f_{\nu\rho]}=0.
\cr}\right.
\equation
$$
The four dimensional spinors $\psi$ are releated to $\Phi$ and $\chi$ as:
$$
\psi=e^{ims}\pmatrix{\Phi\cr\chi}.
\equation
$$
In  Minkowski space, using
light-cone coordinates, the metric is written $d\vx^2+2dtds$, and the
Dirac matrices can be chosen as:
$$
\gamma^t=\pmatrix{0&0\cr1&0},
\qquad
\vec{\gamma}=\pmatrix{-i\vsigma &0\cr0&i\vsigma},
\qquad
\gamma^s=\pmatrix{0&-2\cr0&0}.
\equation
$$
The chirality operator $\Gamma$ is:  
$$
\Gamma\equiv\gamma^5
=
-\smallover{{\sqrt{-g}}}/{4!}\,\epsilon_{\mu\nu\rho\sigma}
\gamma^\mu\gamma^\nu\gamma^\rho\gamma^\sigma,
\equation
$$
in coordinates:
$$
\Gamma=
\pmatrix{-i\sigma_3&0\cr0&i\sigma_3}.
\equation
$$

        In four dimensions both the vector and the axial coupling can be used
in equations (2.6), because the matter field is massless.
Carrying out the light-like Kaluza-Klein type reduction on the equations
(2.6) coupled by the axial current:
$$ 
j^\mu=\overline\psi\gamma^\mu i\Gamma\psi,\qquad 
\overline\psi =\psi^\dagger G,\qquad G=\pmatrix{0&1\cr 1&0}, 
$$
results in our 2+1 dimensional problem. Thus this model is the axial 
type counter part of the model considered in [3]. Also the symmetries of 
the two system are the same: The $\xi$-preserving conformal transformations
$$
L_Xg=k\,g
\qquad
\and
\qquad
L_X\xi=0,
\equation
$$
( where k is smooth function ),
act as symmetries on the axial coupled spinor-CS system (2.1,2,3). To show this
we can use the same process employed in [5], with the
same results for the transformations of the equations (2.6), and we  conclude 
the transformations are symmetries if
$$
\delta_X\left(\overline\psi\gamma^\mu i\Gamma\psi\right)
=
\delta_X\overline\psi\gamma^\mu i\Gamma\psi
+
\overline\psi\gamma^\mu i\Gamma\delta_X\psi
=
L_X\left(\overline\psi\gamma^\mu i\Gamma\psi\right) 
+ 2 k\,\overline\psi\gamma^\mu i\Gamma\psi.
$$
namely if the axial Dirac current transforms in the same way as the CS current 
does.
Since this is so, we find indeed that: the $\xi$-preserving conformal 
transformations are symmetries of the system. These symmetries form a finite 
dimensional Lie group, called the extended `Schr\"odinger' group [6].

        Let us see another version of our model! Realising  that
the lower component of $\psi$ can be expressed from (2.1) simply:
$
\chi=-(1/2m)(\vec{\sigma}\cdot\vec{D})\Phi,
$
we find, that $\Phi$ is the relevant component i.e. one can write a
system of equations including only the $\Phi$. Thus the equations to 
solve are:
$$
iD_t\Psi_\pm=
\Big[-{\vec{D}^2\over2m}
\pm\lambda\,(\Psi_\pm^\dagger\Psi_\pm)\Big]\Psi_\pm,
\qquad
\lambda\equiv{e^2\over2m\kappa},
\equation
$$
together with the FCI.
For the static case they simplify to:
$$\left\{\eqalign{
&\Big[-{1\over2m}(\vec{D}^2+eB\sigma_3)-eA_t\Big]\Phi=0,
\ccr
&\vec{J}=-{\kappa\over e}\vec\nabla\times A_t,
\ccr
&\kappa B=-e\varrho.
\cr}\right.
\equation
$$
I will use mainly this form in what follows.

        We now have a system described by its equations of motion, and we know,
that these equations are self consistent. There is another way to describe 
a model, namely describing it 
with an action. It is an important question whether we can  find an action
producing equations (2.1,2) with the current (2.5). Looking at the 2+1 
dimensional action 
$\int\!d^3x\L$ with
$$\eqalign{
&{\cal L}={\kappa\over4}\epsilon^{\mu\nu\rho}A_\mu F_{\nu\rho}+
\ccr
&\left\{\bigl[
\Phi^\dagger\sigma_3
\big(D_t\Phi+i\sigma^iD_i\chi\big)
\bigr]
+\bigl[
\chi^\dagger\sigma_3
\big(\sigma^iD_i\Phi+2m\chi\big)
\bigr]\right\}
\cr},
\equation
$$
we find that the density and the current, following from this action, 
( the coefficient of $-iA_0$ and $-i\vec{A}$ in the spinor part of $\L$ ), 
is the same, as the current 
in eq. (2.5), and the variational equations are indeed equivalent to (2.1,2) 
(they give the Levi-Leblond equations multiplied by the constant
matrix $\sigma_3$ ).

\chapter{Self-dual solutions}

        Let us consider  the solutions of this model. First ( using the 
similarity of the model to that of in [3]) we try to find time independent 
self dual solutions like the ones found there.
In doing so I assume the same self duality condition wich is used in [3]. 
Notice, that for solutions in wich the two-component spinor,
$\Phi$, has only one (upper or lower) nonvanishing component, $\rho$ has a
definite sign, and is given by $\varrho=|\Psi_+|^2$ or $\varrho=-|\Psi_-|^2$. 
The self duality condition of [3] can be written:
$$
\big(D_1\pm iD_2\big)\Phi=0.
\equation
$$ 
As a result if $\Phi$ satisfies (3.1) we can replace $\vec{D}^2=D_1^2+D_2^2$
by $\mp eB$ in our equations (2.13).
        Let's consider the static self dual version of our system!
Only the first equation of (2.13) changes its form:
$$
\Big[-{1\over2m}eB(\mp 1+\sigma_3)-eA_t\Big]\Phi=0,
\equation
$$
and the others remain as in (2.13).
        One can see, that there are solutions with vanishing $A_t$, and
$\vec{J}$ by choosing $\Phi=\Phi_+$ ( or $\Phi=\Phi_-$ ), where
$$
\Phi_+=\pmatrix{\Psi_+\cr0\hfill\cr}
\and
\Phi_-=\pmatrix{0\hfill\cr\Psi_-\cr}.
\equation
$$
It can be seen easily, that $\chi$ is zero for this choice. 
These expressions solve the static system, if they solve the 
remaining equations of self duality and $B\=-(e\varrho)/\kappa$, where $\rho$ 
has the special form given above.
We get from the self duality:
$$
\vec{A}\=\pm{1\over2e}\vec{\nabla}\times\ln|\Psi_\pm|^2.
\equation
$$
In the gauge $\Psi_+\=\rho^{1/2}$ or ( $\Psi_-\=-\rho^{1/2}$ ) the remaining 
equations reduce to the Liouville equation
$$
\bigtriangleup\ln|\Psi_\pm|^2={2e^2\over\kappa}|\Psi_\pm|^2.
\equation
$$
The $\pm$ sign in the connection between $\rho$ and $\Psi_\pm$ and in the
solution of the self-duality conditions combined nicely to yield a universal
plus sign in (3.5).
Normalizable solutions are obtained when $\kappa<0$.
The well known general solution of the Liouville equation is: 
$
\pm\varrho=|\Psi_\pm|^2=-(4\kappa/e^2)|f'(z)|^2(1+|f(z)|^2)^{-2}
$
where $f(z)$ is complex analytic. 
As we can see these solutions of the axial type
problem are very similar to  the solutions of the vector one. 
It is easy to see the reason behind this: In this very special situation, 
when $\Phi$ has only one nonzero component, our generally indefinit density, 
$\rho$, becomes definite, and differ just in a + or $-$ sign from the 
density in the vector model. We will see however, that these are not the only 
possible static solutions of our model.

\chapter{The new self-duality and the \lq\lq vortex lattice" solutions}

Let's consider how the LL equations change when the self-duality and the 
special form of $\Phi$, ($\Phi_+$ or $\Phi_-$ see in eq (17)) is used. It 
is easy to see, that in this case the LL simplify to:
$$\left\{
\matrix{
2m\,\chi&=&0,
\ccr
D_t\,\Phi&+\hfill&i(\vec{\sigma}\cdot\vec{D})\,\chi\hfill&=&0.
\cr}\right.
\equation
$$
These are the equations wich can be satisfied with the static solutions, having
$\chi=0$ and $A_t=0$. The current is zero because $\chi=0$. The remaining 
equations are 
$B\=-(e\varrho)/\kappa$ plus some conditions, simplifying LL to (4.1).
If we don't want the special one-component form of $\Phi$, but want more
general $\Phi$-s, we must choose the condition, 
what solves the LL equations, in the new form as:
$$
(\vec{\sigma}\cdot\vec{D})\,\Phi=0.
\equation
$$
This is our new self-duality condition\foot{ This kind of self duality equation
can be used to solve the vector model as well. This condition
minimalizes the effective Hamiltonian for $\Phi$ in the vector model too.}.
For $\Phi$-s having only upper ( or lower) nonzero components it 
requires the same as the earlier condition (3.1).
        What are the equations we have to solve now? As we have seen, static 
solutions with $\chi=0$ and $A_t=0$ satisfy the LL equations with the new 
self-duality. For this kind of solutions 
$\kappa E^i\equiv -\partial_iA^0-\partial_tA^i=
e\,\epsilon^{ij}J^j
$
is an identity.
The remaining equations that must be solved, are the new self-duality 
condition (4.2), and the `CS Gauss law':
$$
B\=-(e/\kappa)\varrho,
$$
where $\varrho=|\Psi_+|^2-|\Psi_-|^2$  in this model.
In the gauge $\Psi_+=|\Psi_+|$, $\Psi_-=|\Psi_-|e^{i\alpha}$ we find 
from them:
$$
\vec{A}=
{1\over2e}\vec{\nabla}\times\ln|\Psi_+|^2\quad 
=-{1\over2e}\vec{\nabla}\times\ln|\Psi_-|^2+{1\over e}\vec{\nabla}\alpha.
\equation
$$
Therefore $\vec{\nabla}\times\ln|\Psi_+|^2=-\vec{\nabla}
\times\ln|\Psi_-|^2+2\vec{\nabla}\alpha$. 
 Using $\vec{\nabla}\times\vec{\nabla}\alpha =0$ it follows, that 
$\bigtriangleup\ln|\Psi_-|^2|\Psi_+|^2=0$,  
thus  
$
\ln|\Psi_-|^2|\Psi_+|^2=4F(x,y)$, where $F$ is real and harmonic,
that is $\bigtriangleup F=0$. From the definition of $F$ we get
$$
|\Psi_-|^2=e^{4F}|\Psi_+|^{-2}.
$$
Using the Chern-Simons Gauss law yields 
$$
\bigtriangleup\ln|\Psi_+|^2
=
-{2e^2\over\kappa}\big(|\Psi_+|^2-e^{4F}|\Psi_+|^{-2}\big).
\equation
$$
It is useful to work with the complex variable $z=x+iy$. If $z=f(w)$ is an 
analitic function, then $\bigtriangleup_w=|f'|^2\bigtriangleup_z$.
Introducing 
$$
\sigma=\ln|\Phi_+(f(w))|^2+\ln|f'(w)|^2,$$ equation (4.4) can be simplified
further as 
$$
\bigtriangleup_w\sigma=-\big({2e^2\over\kappa}\big)
\big(e^\sigma-(e^F|f'|)^4e^{-\sigma}\big).
$$
Choosing $f$ so that $|f'|=e^{-F}$ we get from here:
$$
\bigtriangleup\sigma=-\big({4e^2\over\kappa}\big)
\sinh(\sigma).
\equation
$$
This equation shows a manifest scaling symmetry, and after
the scale transformation $w'={2e\over\kappa^{1/2}}w$, the constant $4e^2\over
\kappa$ is scaled to 1 and we have:
$$
\bigtriangleup\sigma=-\sinh(\sigma).
\equation
$$
This is the two dimensional sinh-Poisson equation, wich has well known  
numerical [7,8], and analitical [9] solutions such as the solutions of a non 
linear  boundary value problem on a square or on a rectangle. In these 
papers some arguments are put forward which indicate, that there are no 
regular solutions on the whole plane, just on finite regions. 
The large structure forms a vortex lattice on the plane. The boundary 
conditions in x and y, releated to periodic solutions are: $\sigma=0$ on the 
sides of a square. The functions solving eq. (4.6) with this 
boundary conditions are:
$$
\sigma=2\ln\Big[{\theta(\vec{l}+1/2\vec{1},\tau)\over\theta(\vec{l},\tau)}\Big],
\equation
$$
where 
$$
\theta(\vec{l},\tau)=\sum_{m_1..m_N=-\infty}^\infty\exp(2i\pi\sum_{i=1}^Nm_il_i
+i\pi\sum_{i,j=1}^Nm_i\tau_{ij}m_j),
$$
is the Riemann theta function,
$$
l_j(x,y)=k_jx+\omega_jy+l_{0j},\qquad 
\vec{1}=(1,...,1),
$$
and $k_j$, $\omega_j$, $\tau_{ij}$ have nontrivial dependence on a set of 2N 
complex parameters $E_j$-s ( called the main spectrum ) through 
some closed contour integrals on a two-sheeted Riemann-surface. 
These contours run through and around the branch cuts begining 
and ending at the $E_j$-s. The structure of the main spectrum determines the 
properties of $\sigma$. If the $E_j$-s have some special symmetries $\sigma$ 
becomes real. These real solutions are non linear "standing waves", and their 
periods  depend on the wave numbers $k_j$ and $\omega_j$.
        Returning to our physical problem, for the interesting quantities, 
$\rho$ and $\vec{A}$, we have the following results:
$$
\rho=e^{2F}\sinh(\sigma)=e^{2F}\bigg[{\theta(\vec{l}+1/2\vec{1},\tau)^2\over
\theta(\vec{l},\tau)^2}-{\theta(\vec{l},\tau)^2\over\theta(\vec{l}+1/2\vec{1},
\tau)^2}\bigg],
\equation
$$
in terms of the scaled variables. If we want periodicity and smooth transition 
from domain to domain, we must choose $\exp(2F)$ to be a constant, denote it 
by $c$. This means the following: the phase shift, $\exp(i\alpha)$, between 
$\Psi_+$ and $\Psi_-$ is an x and y independent quantity. 
( $\vec{\nabla}\alpha=\vec{\nabla}\times F=0$ ). After it we can write:
$$
\rho=c\bigg[{\theta(\vec{l}+1/2\vec{1},\tau)^2\over\theta(
\vec{l},\tau)^2}-{\theta(\vec{l},\tau)^2\over\theta(\vec{l}+1/2\vec{1},
\tau)^2}\bigg]\ ,
\equation
$$
and in a general gauge:
$$
\Phi=\pmatrix{c^{1/2}{\theta(\vec{l}+1/2\vec{1},\tau)\over\theta(\vec{l},\tau)}
\cr
c^{1/2}e^{i\alpha}{\theta(\vec{l},\tau)\over\theta(\vec{l}+1/2\vec{1},\tau)}
\cr}
e^{i\beta(x)},
\equation
$$
$$
\vec{A}={1\over2e}\nabla\times\ln\bigg[{\theta(\vec{l}+1/2\vec{1},\tau)\over
\theta(\vec{l},\tau)}\bigg]+{1\over e}\nabla\beta(x).
\equation
$$
The value of the constant $c$ is quantised by the boundary condition, 
and, what can be seen easily,
it depends on the number density of particles with + or - spin polarity.

The stability of the solutions can be studied [7] with numerical methods, and
an important result of these numerical calculations is, that the solution  
having the greatest size, (in wich the vortex number per domain is the 
smallest possible), 
is the thermodinamically stable one.
\vfill\eject
\chapter{Chiral coupling}

Notice, that we already proved the possibility and consistency of the 
vector and axial vector couplings. This suggests that there can be 
consistent
systems coupled by the ( left or right ) chiral currents. The chiral invariance
of the four dimensional Dirac-equation is well known. It can be seen easily,
that the four dimensional system, (2.6), with the current
$$
j_\mu={1\over2}(j^v_\mu\pm j^{ax}_\mu)={1\over2}\overline\psi
\gamma_\mu (1\pm i\Gamma)\psi,
\equation
$$
is well defined. The $\xi$-preserving conformal transformations are symmetries
of these systems too, following from the results above and from the linearity 
of the transformations.

The 2+1 dimensional equations which one obtaines after the reduction are 
equations (2.1) and (2.2) with the current $\vec{J}_{L,R}={1\over2}(\vec{J}^v\pm
\vec{J}^{ax})$, where $\vec{J}^v$ is the 2+1 dimensional vector and 
$\vec{J}^{ax}$ is the axialvector current (defined in (2.4) and (2.5)).
$$
\rho_{L,R}=|\Psi_\pm|^2,
\equation
$$
$$
J_{L,R}^1=-(\psi^\dagger_\pm\sigma^2\psi_\pm ),
$$
$$
J_{L,R}^2=\pm (\psi^\dagger_\pm\sigma^1\psi_\pm ),
$$
where
$$
\psi_+=\pmatrix{\Psi_+\cr\chi_+\cr}
\and
\psi_-=\pmatrix{\Psi_-\cr\chi_-\cr},
\equation
$$
are the chiral components of the s-independent part of the Dirac spinor $\psi$.

Note: $J^1_L$ and$J^2_L$ are related to the imaginary and real parts of the
complex function $\Psi^*_+\chi_+$ and we have a similar relation for $\vec{J}_R$
and $\Psi^*_-\chi_-$.

Using the linearity we can see, that the continuity equation is valid for the
density $\rho_{L,R}$ and the current 
$\vec{J}_{L,R}$, and we can look for the static solutions of this consistent
system of equations. The LL equations split into two uncoupled systems for the
chiral components, but both of these have static $A_t=0$ solutions, with 
$\vec{J}_{L,R}=0$, if the self-duality (4.2) is valid. Notice we must 
solve now the same self-dual system, what we have solved in sec. 2., and after 
some simple calculations we end up again with the Liouville equation. ( The 
chiral densities have the special form (3.3). The remaining equations with this
kind of $\rho$ lead to the Liouville equation, as we have seen it earlier. ) 
There are normalisable solutions for the left chiral case when $\kappa<0$, and 
for the right chiral coupling when $\kappa>0$, thus the sign of $\kappa$ 
has to be connected with the sign determining the chirality.

\chapter{Disscusion}

In a model in wich the Chern-Simons gauge field is coupled to a spinor field,
not only the vector current can play the role of the coupling current, but the
axial and chiral type currents are possible too. In this paper we studied the
other possibilities, and looked for static self-dual solutions.
With the axial type coupling I found a kind of vortex 
lattice in wich alternating regions are dominated by up and down spin states.
In the chiral case the problem simplified to the solution of the Liouville 
equation, what is well known. The required sign of the CS coupling constant
$\kappa$ is connected with the "chirality".

\vskip 1cm

I would like to thank L. Palla for calling my attention to the 
spinor- Chen-Simons theory, and for many helpful discussions.

\vskip 1cm


\centerline{\bf\BBig References}

\bigskip

\reference
R.~Jackiw and S.-Y.~Pi,
Phys. Rev. Lett. {\bf 64}, 2969 (1990);
Phys. Rev. {\bf 42}, 3500 (1990);
see Prog. Theor. Phys. Suppl. {\bf 107}, 1 (1992) for a review.

\reference
P.~Olesen,
Phys. Rev. {\bf B265}, 361 (1991).

\reference
C.~Duval, P.~A.~Horv\'athy and L.~Palla, 
Phys. Rev. {\bf D52}, 4700 (1995).

\reference
J.-M.~L\'evy-Leblond, 
Comm. Math. Phys. {\bf 6}, 286 (1967).

\reference
C.~Duval, P.~A.~Horv\'athy and L.~Palla,
Ann. of Phys. {\bf 249}, 256 (1996).

\reference
R.~Jackiw, 
Phys. Today {\bf 25}, 23 (1972);
U.~Niederer, 
Helv. Phys. Acta {\bf 45}, 802 (1972);
C.~R.~Hagen, 
Phys. Rev. {\bf D5}, 377 (1972);
G.~Burdet and M.~Perrin, 
Lett. Nuovo Cim. {\bf 4}, 651 (1972).

\reference
Y.~B.~Pointin, T.~S.~Lundgren,
Phys. Fluids {\bf 19}, 1459 (1976).

\reference
D.~L.~Book, S.~Fisher, and B.~E.~McDonald,
Phys. Rev. Lett. {\bf 34}, 4 (1975).

\reference
A.~C.~Ting, H.~H.~Chen, Y.~C.~Lee,
Phys. Rev. Lett. {\bf 53}, 1348 (1984);
A.~C.~Ting, at al.,
Physica {\bf 26D}, 37 (1987)

\bye